\documentclass{article}
\usepackage[T1]{fontenc}
\usepackage{graphics}

\makeatletter

\newcommand{\LyX}{L\kern-.1667em\lower.25em\hbox{Y}\kern-.125emX\@}

\makeatother

\begin{document}

\title{\textbf{Tools for RHIC: Review of Models}\thanks{
Invited talk at the 5th International Conference on Strangeness in Quark Matter
}}

\author{\textbf{K. Werner} {\small }\\
{\small }\\
{\small Laboratoire de Physique Subatomique et des Technologies Associ\'{e}es
(SUBATECH)}\\
{\small Universit\'{e} de Nantes, IN2P3/CNRS, Ecole des Mines de Nantes}\\
 {\small 4, rue Alfred Kastler, F-44070 Nantes Cedex 03, France.}\small }

\maketitle
\begin{abstract}
We discuss the present status of microscopic models for RHIC, with an emphasis
on models being realized via the Monte Carlo technique. This review is to a
large extent based on the OSCAR3 workshop, where general concepts and new trends
in this field have been discussed. 
\end{abstract}

\section{Introduction}

Although very interesting data have been collected during the SPS program of
heavy ion physics, no clear quantitative conclusions can be drawn concerning
the formation (or not) of a quark gluon plasma. A well established theory exists
(QCD), however, technical difficulties prevent a direct application of the theory
to understand data. Effective theories have been proposed to overcome these
difficulties, as well as very simple qualitative models, which might as best
be called ``theory inspired''. On the other hand, the so-called ``event generators''
or ``Monte-Carlo codes'' have been introduced, which by definition provide
randomly generated ``events'', characterized by a certain number of particles
of different types with given momenta. One aim of this paper is to discuss general
strategies how such event generators should be constructed in order to provide
useful tools to understand experimental data.
\begin{figure}[htb]
{\par\centering \resizebox*{!}{0.3\textheight}{\includegraphics{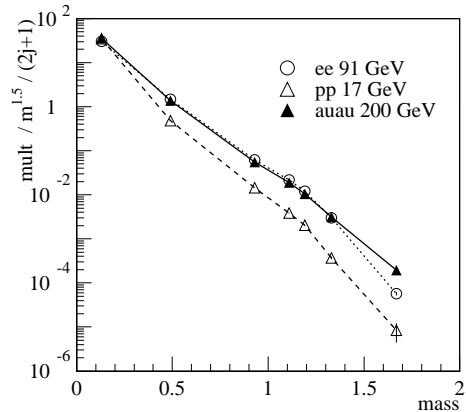}} \par}

\caption{Hadron yields as a function of the hadron mass for different reactions: electron-positron
annihilation at 91 GeV (open dots), pp scattering at 17 GeV (open triangles),
and AuAu scattering at 200 GeV(full triangles).\label{fig:yields}}
\end{figure}

There is some effort to be done, but it is worth it, because finally MC codes
are absolutely necessary to understand data. As an illustration, we plot in
fig. \ref{fig:yields} yields of different hadrons as a function of the hadron
mass for different reactions: electron-positron annihilation at 91 GeV, pp scattering
at 17 GeV, and AuAu scattering at 200 GeV. The curves are arbitrarily normalized.
We observe a very interesting and unexpected result: all the spectra are roughly
exponential, and the spectra for heavy ion collisions agree with the result
for electron-positron. Here, all the results are based on calculations, and
so we know that the dynamics of electron-positron annihilation is completely
different to the one in heavy ion collisions. So one cannot draw any conclusion
concerning a plasma formation based on the fact that strange particles and in
particular (multi-) strange baryon production is enhanced, since the same effect
may be due to a completely different mechanism (as in electron-positron annihilation).

\section{OSCAR Philosophy}

Monte-Carlo simulations are often criticized of not being well documented, having
no clear physical basis, being constantly modified, being not publicly available,
and so on, which makes all evaluation of the quality of such an approach impossible.
In order to improve the situation, OSCAR was founded a couple of years ago,
which is first of all a series of workshops, hold every one or two years, as
well as working groups and a permanently updated WEB page\footnote{
http://nt3.phys.columbia.edu/OSCAR
}. In the working groups, mainly technical issues have been discussed, like standards
for input/output, test for individual moduls, and so on. The first two workshops
were held in Brookhaven in 97 and 99, organized by Y. Pang and M. Gyulassy,
the third one took place in Nantes in 2000, organized by Y. Schutz and K. Werner. 

At the first two workshop the general philosophy was still based on the fact
that MC models can hardly be defined by equations, they were mostly defined
in a algorithmic way, and therefore the only way to control these codes is good
documentation, definition of standards, accessibility, modular structure, all
this in particular to allow intensive testing of these codes. 

Starting with the third workshop a new philosophy has been discussed, in an
attempt to make the event generators much more transparent, and thus to provide
really useful tools for analyzing data. So the model should be simply defined
by equations, and the computer code should be just the technical means to solve
these equations, briefly:

\begin{quote}
{\par\centering model = equations,\par}

{\par\centering MC code = solution of the equations.\par}
\end{quote}
It is clear that these equations building the basis of the MC code will (probably)
never be directly derived from first principles, this is also not necessary.
One may construct some effective theory which may be simply inspired by the
``true'' theory, but the next step, the MC implementation, has to rigorous.
The advantage is obvious: is is difficult to have any meaningful discussion
about some code where the physical basis is not well defined. When the code
however is nothing but the numerical treatment of some very well defined (even
if not fundamental) effective theory, one is at least able to have some physics
discussion, develop further promising developments, and eliminate those ones
based on wrong physical ideas. There was very little progress in this direction
during the past years.

\section{The Different Stages of Heavy Ions Collisions}

Unfortunately there does not exist a single formalism being able to account
for a complete nucleus-nucleus collision. Rather we have to -- at least for
the moment -- to divide the reaction into different stages
\begin{figure}[htb]
{\par\centering \vspace{-1cm}\par \resizebox*{!}{0.35\textheight}{\rotatebox{270}{\includegraphics{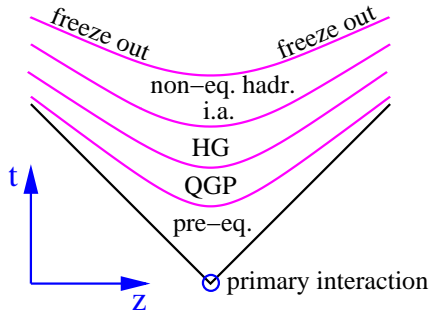}}} \\
\vspace{-1cm}\par\par}

\caption{The different stages of heavy ion collisions.\label{stages}}
\end{figure}
(see fig. \ref{stages}) and try to understand the different stages as good
as possible. 

There is first of all the primary interaction when the two nuclei pass through
each other. Since at very high energies the longitudinal size is due the gamma
factor almost zero (of the order 0.1 fm at RHIC), all the nucleons of the projectile
interact with all the nucleons of the target instantaneously. In such a primary
interaction many partons are created, which interact (in the pre-equilibrium
stage) before reaching an equilibrium, referred to as quark-gluon plasma. The
system the expands, passing via phase transition (or sudden crossover) into
the hadron gas stage. The density decreases further till the collision rate
is no longer large enough to maintain chemical equilibrium, but there are still
hadronic interactions till finally the particles ``freeze out'', i.e. they
continue their way without further interactions.
\begin{figure}[htb]
{\par\centering \vspace{-1.5cm}\par\resizebox*{!}{0.45\textheight}{\rotatebox{270}{\includegraphics{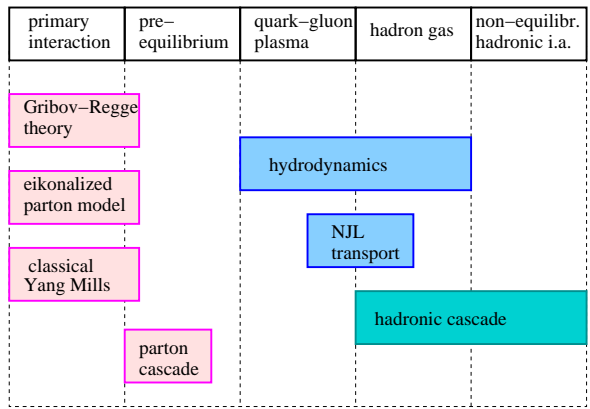}}} \\
\vspace{-1.5cm}\par\par}

\caption{The different stages of heavy ion collisions and the corresponding range of
validity of theoretical approaches.\label{models}}
\end{figure}
Different theoretical approaches have been proposed, which are valid only for
certain stages of the collision, as indicated in fig. \ref{models}. We will
present the details in the following sections. This discussion will certainly
be quite incomplete, with few exceptions we restrict ourselves to topics having
been discussed at OSCAR 3.

\section{The primary interaction}

\subsection{Longitudinal Excitation}

Let us start the discussion with a approach which is widely used today to simulate
the primary interactions, but which is definitely wrong: the longitudinal excitation.
At low energy nucleon-nucleon scattering, a typical reaction is the excitation
of nucleon resonances via the exchange of a meson. 
\begin{figure}[htb]
{\par\centering \vspace{-1.5cm}\par\resizebox*{!}{0.25\textheight}{\rotatebox{270}{\includegraphics{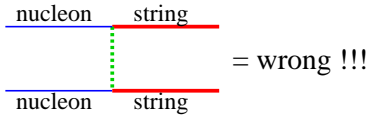}}} \\
\vspace{-1.5cm}\par\par}

\caption{Longitudinal excitation (which is kinematically impossible).\label{long}}
\end{figure}
So one might me tempted to generalize this mechanism to high energies: two nucleons
interact via the exchange of ``something'' which causes the two nucleons to
be excited to strings, the latter ones considered to be the high energy generalization
of excited nucleon states. However, such an extrapolation from low to high energies
is simply wrong, due to kinematical reasons. Let us consider two-body kinematics:
two incoming particles with four-momenta \( p_{1} \) and \( p_{2} \) interact
in an non-specified way and produce two outgoing particles with momenta \( p_{3} \)
and \( p_{4} \). The transferred momentum is defined to be \( q=p_{1}-p_{3} \).
A short calculation shows:
\[
q^{2}=q_{\bot }^{2}+O(\frac{1}{s}),\]
 where \( q_{\bot } \) is the transverse component of the transferred momentum,
orthogonal to \( p_{1} \) and \( p_{2} \). At high energies (\( s\gg 1 \)
GeV\( ^{2} \)), the transferred momentum is consequently purely transverse,
there is no transfer of longitudinal momentum. String excitation, on the other
hand, requires a transfer of longitudinal momentum in order to allow for string
with non-zero mass.

All this has been known since forty years, which does not prevent people from
coming up with models based on this wrong idea, or using such models in trying
to understand data.

\subsection{Yang-Mills Equations}

For very large nuclei and correspondingly high parton densities screening will
be most effective, and therefore soft physics can be completely accounted for
by assuming random color sources moving along the light cones, the latter ones
generating chromoelectric fields calculable by solving the corresponding classical
Yang-Mills equations \cite{mcl94}. 

This is an interesting theoretical idea, although it is not clear how to construct
an ``event generator'' based on such an approach, so we do not want to discuss
any details here.

\subsection{The Parton Model}

The parton model approach to nucleon-nucleon scattering amounts to presenting
the partons of projectile and target by momentum distribution functions, \( f_{i} \)
and \( f_{j} \), and calculating inclusive cross sections for the production
of parton jets as a convolution of these distribution functions with the elementary
parton-parton cross section \( d\hat{\sigma }_{ij}/dp_{\perp }^{2} \), where
\( i,j \) represent parton flavors. 

This simple factorization formula is the result of cancelations of complicated
diagrams (AGK cancelations) and hides therefore the complicated multiple scattering
structure of the reaction, which is finally recovered via eikonalization procedure.
The latter one makes the approach formally equivalent to the Gribov-Regge one,
to be discussed later. Generating events and particle production is not at all
evident in this approach. The Pythia-method \cite{sjo87} amounts to generating
the first elementary interaction according to the inclusive differential cross
section, then taking the remaining energy for the second one and so on. In this
way, the event generation will reproduce the theoretical inclusive spectrum
for hadron-hadron interaction (by construction).

Concerning nucleus-nucleus collisions, one usually assumes the proton-proton
cross section for each individual nucleon-nucleon pair of a \( AB \) system.
Nuclear screening effects may be taken into account by using \( A \)-dependent
parton distribution functions, \( f^{A}_{i} \) and \( f^{A}_{j} \), rather
than the ones used for nucleon-nucleon scattering (this is usually referred
to as shadowing).

The HIJING model \cite{wan96} is constructed along these lines, with the additional
feature of considering the energy loss of partons due to final state interactions.

\subsection{Gribov-Regge Theory}

\begin{figure}[htb]
{\par\centering \vspace{-2cm}\par\resizebox*{!}{0.4\textheight}{\rotatebox{270}{\includegraphics{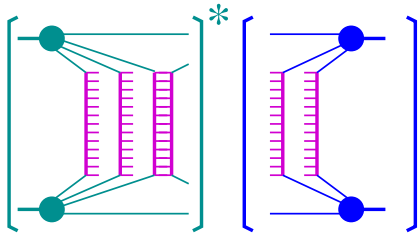}}} \\
\vspace{-2cm}\par\par}

\caption{An interference term in GRT.\label{grt}}
\end{figure}

Gribov-Regge theory (GRT) has been developed well before QCD, but it is even
today more relevant than ever, with HERA giving the opportunity to test and
verify the different aspects of this approach. There is no strict derivation
from first principles, one can just follow some ``QCD inspired arguments''
to write down an expression for the elastic scattering amplitude for nucleon-nucleon
scattering in terms of many elementary scatterings, which easily generalizes
to nucleus-nucleus scattering. From there on, one uses strictly the rules of
quantum mechanics to obtain a multiple scattering approach for particle production.
The key ingredient is the fact that the cross section is obtained from squaring
the amplitude, so one obtains partial contributions as shown in fig.\ref{grt}.
 In the example shown in the figure, we have a diagram with two inelastic interactions
(symbolically shown as ``comb'') interfering with the diagram with two inelastic
and one elastic scattering (symbolically shown as ``ladder''). So we have
classes of interfering contributions, which have to be summed up. Ignoring energy
conservation, one obtains a simple formula for the inelastic cross section:
\begin{equation}
\label{inel}
\sigma _{\mathrm{inel}}(s)=\int d^{2}b\sum _{m}\frac{f(s,b)^{m}}{m!}e^{-f(s,b)}.
\end{equation}
 The function \( f \) is related to the inclusive cross section as 
\begin{equation}
\label{incl}
f(s,b)=\sigma _{\mathrm{incl}}(s)\, A(b),
\end{equation}
with some function \( A \) representing the impact parameter dependence. The
equations (\ref{inel}, \ref{incl}) provide the link between the parton model
and GRT: since the parton model only provides inclusive cross sections, one
uses these formulas to introduce multiple scattering. Or one may say it the
other way round: GRT models introduce hard scattering via inclusive cross sections
calculated in the parton model.

Models based on the Gribov-Regge approach are QGS \cite{kai82}, DPM \cite{cap94,aur94},
VENUS \cite{wer93}.

\subsection{String Fusion}

Colliding heavy nuclei at very high energies will produce a large number of
strings, which can certainly not be considered completely independent as in
standard GRT. A phenomenological approach consists of considering string fusion,
whenever pairs of strings are coming close to each other \cite{arm97}. Fused
strings are assumed to behave like ordinary strings simply with a increased
string tension. This will, for example, lead to an enhanced production of multi-strange
particles.

\subsection{Nuclear Shadowing}

There exists also a formally well defined treatment of high density effects
via the so-called enhanced diagrams, which amounts to a generalization of the
basic Gribov-Regge theory by taking into account Pomeron-Pomeron interactions.
This is responsible not only for nuclear shadowing, but also for diffractive
scattering in \( pp \) but also in deep inelastic scattering. In fact it is
very important to have a consistent picture of \( pp \) and DIS, in general
and in particular for diffractive scattering \cite{van00}. The latter one reveals
a reduction of the Pomeron flux compared to a naive expectation from GRT, which
then leads to a reduced screening. To say it differently, one has to go beyond
simple lowest order enhanced diagrams (triple Pomerons).

\subsection{Proper Energy Sharing in GRT}

Although GRT is a useful starting point to describe high energy hadronic and
nuclear scattering, there are serious drawbacks. As pointed out in \cite{dre00},
GRT is lacking a consistent picture for the calculation of the cross section
formulas and for particle production. The problem is the energy sharing between
the individual scatterings. Doing this properly makes the approach considerably
more complicated, and therefore the standard approach is to ignore energy sharing
at this level, and considering it later when it comes to particle production.
This is clearly not consistent, and in fact the error due to neglecting the
energy sharing is quite large: the width for the distribution of the number
of multiple scatterings is roughly doubled when energy conservation is ignored.
This can be compensated by making a second mistake which amounts to ignoring
the increase of multiplicity fluctuations due to cutting enhanced diagrams properly.
A new model \textsc{\large neXus} has been proposed recently\cite{dre00}, where
energy conservation is treated properly on all levels. In addition, lowest order
enhanced diagrams are considered. Soft and hard scattering is treated such that
there is a smooth transition between the two regimes, and in particular a dependence
on some transverse momentum cutoff \( p_{0} \) can be avoided.

\section{Pre-Equilibrium}

The partons created in the primary interactions are certainly far from equilibrium,
and is desirable to understand microscopically the equilibration of the system,
in other words the formation of a quark gluon plasma. This is a difficult task,
since for example at RHIC energies there is still a large soft component. Nevertheless
it is useful to study the evolution of partonic systems based on pQCD, ignoring
soft physics.

\subsection{Parton cascade}

A parton cascade amounts to considering partons as classical particles which
move on straight line trajectories, where binary interactions are defined via
parton-parton cross sections calculated in the framework of perturbative QCD
\cite{gei92}. 

One has to carefully regard the range of validity of this approach: it is not
meant to treat the primary interactions, where quantum mechanical interference
should play a crucial role, so one may start the calculation once a system of
incoherent classical partons have been established. On the other end, one should
not stretch the perturbative treatment too far: perturbative calculations require
large momentum transfer which is not any more guaranteed if the interaction
energy is getting too low.

A parton cascade is often referred to as the solution of a Boltzmann equation.
In this case one has to work with test particles rather than real particles
and one has to make sure that the number of test particles is sufficient to
really provide a solution of the equation \cite{mol00} (which is in accordance
with the OSCAR principle ``code = solution of equation''). It turns out that
the number of test particles has to be much larger than the number of real particles,
which prevents a cascade of real partons to be considered as a solution of a
Boltzmann equation. Other than this ``particle subdivision test'' other tests
like ``box tests'' should be performed to make sure than a cascade algorithm
solves really a transport equation.

\subsection{Parton equilibration}

There are also analytical approaches attempting to understand parton equilibration
\cite{dum00}. Starting from a parton density given by the parton model with
some cutoff \( p_{0} \), one calculated the evolution of the system based on
a transport equation of the type
\[
p\partial f=-\frac{pu}{\tau }(f-f_{\mathrm{eq}}),\]
 with \( \tau  \) being the relaxation time, which is given as \( \tau =1/\sigma n \),
with \( \sigma  \) being the in-medium cross section, and \( n \) the parton
density. Comparing RHIC and LHC results, the density will be of course bigger
for LHC, and in both cases decreasing with time. Due to the larger screening,
the cross section will be smaller for LHC compared to RHIC, and in both cases
increasing with time. Taking all together one obtains the time dependence of
the relaxation time which peaks around 1.5 fm, with the LHC value being somewhat
bigger than the RHIC one. One finds a free streaming up to around 1.5 fm, then
the equilibration starts.

\subsection{Parton Energy Loss}

Partons created in hard primary interactions will lose energy when traversing
matter, the latter one being projectile or target nuclei or quark matter \cite{wan95,bai97}.
The formulas for the energy loss obtained so far where obtained in the limit
of either large or small system sizes (\( L \)). New developments have been
reported \cite{gyu00}, which allow a calculation of the energy loss for arbitrary
\( L \), by using a systematic expansion in opacity (\( L/\lambda  \)), where
\( \lambda  \) is the mean free path of the parton. It turns out that the expansion
converges rapidly, with the second order already being a very small correction
compared to the first order one.

\section{Equilibrium and Post-Equilibrium}

We are now discussing the final stage of the collision, consisting of QGP phase,
the hadron gas phase, and the very final stage where the hadrons still interact,
but they do not form an equilibrated system any more. We do not treat these
three stages individually, because the models to be discussed latter treat usually
more than just one stage.

\subsection{Hydrodynamics}

The final aim of all the efforts in the field of ultra-relativistic heavy ion
collisions is the creation of a thermalized system of quarks and gluons. Provided
such an equilibrium has been established, one may use hydrodynamics, which is
a macroscopic approach based on energy-momentum conservation and local thermal
equilibrium. Hydrodynamical calculations have been used since a long time, either
assuming particular symmetries and using analytical methods \cite{bay84}, or
full 3-dimensional calculations numerical calculations \cite{ris98}. Recently
a new technique has been proposed, the so-called smoothed particle hydrodynamics
\cite{agu00}, where fields \( \rho (x) \) are represented by particles as
\( \rho _{P}(x)=\sum _{b}\nu _{b}\delta (x-x_{b}) \), and then smoothed:
\[
\rho (x)\rightarrow \rho _{SP}(x)=\int \rho _{P}(x)W(x-x')dx'=\sum _{b}\nu _{b}W(x-x_{b}),\]
with some smoothing kernel \( W \). The advantage is that the hydrodynamical
equations are transformed into a system of ordinary differential equations,
which can be solved by applying standard methods. In this way one may perform
3-dimensional calculations much faster than with traditional methods.

\subsection{Hadronization}

There are several attempts to treat at least the region around the phase transition
in a microscopic way. A possibility is to apply transport theory based on the
\textbf{NJL model} \cite{reh98}, which is an effective theory with a point-like
interaction between two quarks (gluons are not considered explicitly). The model
allows also for hadron production like quark plus anti-quark goes into meson
plus meson. The dynamics is crucially affected by the density and temperature
dependence of quark and hadron masses, one observes for example the formation
of droplets of quark matter rather than homogeneous matter of lower density,
since the latter one would imply higher quark masses.

A completely different hadronization scenario has been proposed based on the
\textbf{confinement} mechanism \cite{hof99}, again ignoring gluons. Quarks
are considered to be classical particles, their dynamics being determined by
a classical Hamiltonian. The latter one contains a string potential and color
factors which force the quarks to form resonances, which subsequently decay
into hadrons.

Another alternative approach is the hadronization via \textbf{coalescence} \cite{csi99}.
Again, starting from a quark-anti-quark plasma, hadronic resonances are formed
based on coalescence, with a subsequent decay into hadrons.

\subsection{Hadronic Transport Theory}

Once a purely hadronic system has been established, a microscopic treatment
based on binary hadronic interactions is feasible. Here, hadrons propagate on
classical trajectories and interact according to hadron-hadron scattering cross
sections. If possible, parameterizations of measured cross sections are used.
A couple of models have been constructed along these lines, like UrQMD \cite{bas98,ble99},
ART \cite{li95}, JAM \cite{nar97}. Unfortunately, not all the necessary cross
sections have been measured to a sufficient precision, and correspondingly the
above-mentioned approaches differ by using different model assumptions for the
cross sections. We emphasize again that hadronic transport codes are a useful
tool to treat the final stage of a heavy ion collision, but not for the primary
interaction.

\section{Outlook}

Historically, most models have been first developed for a certain aspect of
the collision, and have later been extended to include more and more features
in order to give a complete description of the collision. This is not a good
way to proceed, since in this way the models have ``strong parts'' and ``weak
parts'', and the whole approach is not very reliable. A better way is to consider
``moduls'' , describing just one aspect of the collision in the most realistic
fashion, and combine such moduls. Examples, which have been discussed at the
OSCAR3 workshop, are a combination of hydrodynamics and a hadronic cascade (hydro+UrQMD
\cite{bas00}) or a combination of a primary interaction model and hydrodynamics
(\textsc{neXus+SPheRIO}\cite{agu00}). 

\bibliographystyle{pr2}
\bibliography{a}

\end{document}